\documentclass{PoS}

\title{Latest results from the NA61/SHINE beam energy scan with p+p and Be+Be collisions}

\ShortTitle{Latest results from the NA61/SHINE beam energy scan with p+p and Be+Be collisions}

\author{\speaker{Maja Mackowiak-Pawlowska for the NA61/SHINE Collaboration}%
        Warsaw University of Technology (PL)\\
        E-mail: \email{majam@if.pw.edu.pl}}

\abstract{The NA61/SHINE experiment aims to discover the critical point of strongly interacting matter and study the properties of the onset of deconfinement by measurements of hadron production properties in proton-proton, proton-nucleus and nucleus-nucleus interactions in the CERN SPS energy range.

This contribution presents results on the energy dependence of hadron spectra and yields as well as on fluctuations and two-particle correlations in p+p and centrality selected Be+Be collisions. In particular, the energy dependence of the signal of deconfinement, the "horn", observed in central Pb+Pb collisions is compared with the corresponding results from p+p interactions. Also string-hadronic models are tested using hadron spectra and fluctuations measured in p+p interactions. Results on fluctuations (multiplicity and transverse momentum) are presented as a function of the collision energy for Be+Be and p+p collisions in search for the critical point of strongly interacting matter.}

\FullConference{The European Physical Society Conference on High Energy Physics\\
		22--29 July 2015\\
		Vienna, Austria}

\begin{document}

\section{Introduction}
It is a well established fact that matter exists in different states. For strongly interacting matter at least three states are expected: nuclear matter, hadron gas (HG) and a system of deconfined quarks and gluons (often called the quark-gluon plasma - QGP). It is believed that the Universe during its early stage consisted of QGP.  

One of the most important goals of high-energy heavy-ion collisions is to establish the phase diagram of strongly interacting matter by finding the possible phase boundaries and critical points. In principle,  we want to produce the quark-gluon plasma and analyze its properties and the transition between QGP and HG. The NA61/SHINE~\cite{Antoniou:2006mh} experiment located at the north area of the CERN SPS is a fixed target experiment. Its main detectors are four large volume Time Projection Chambers (two of them in super-conducting magnets) and two Time-of-Flight (ToF) walls. Details on the detector can be found in Ref.~\cite{Abgrall:2014xwa}. 

\begin{figure}[h]
\includegraphics[width=0.45\textwidth]{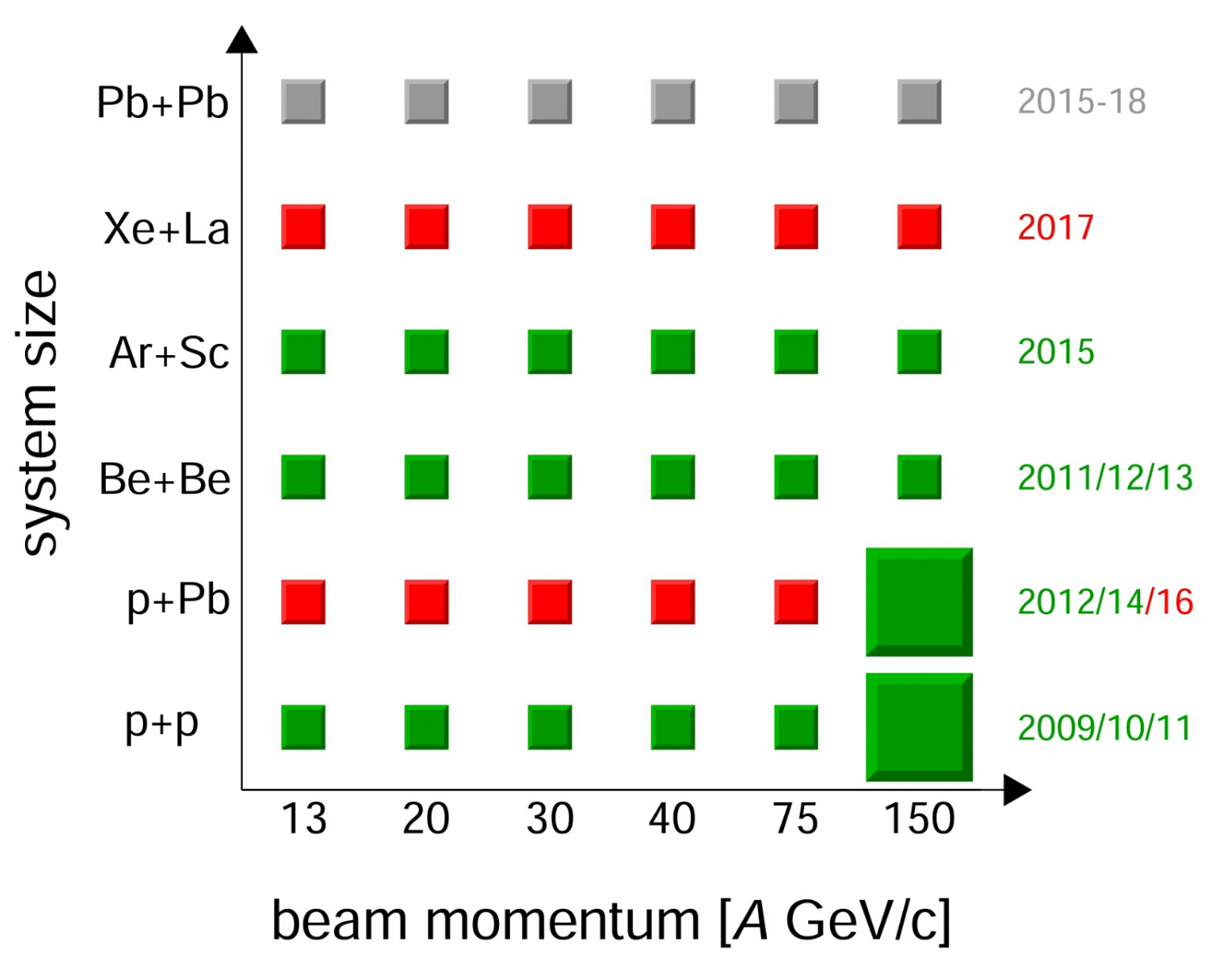}\includegraphics[width=0.45\textwidth]{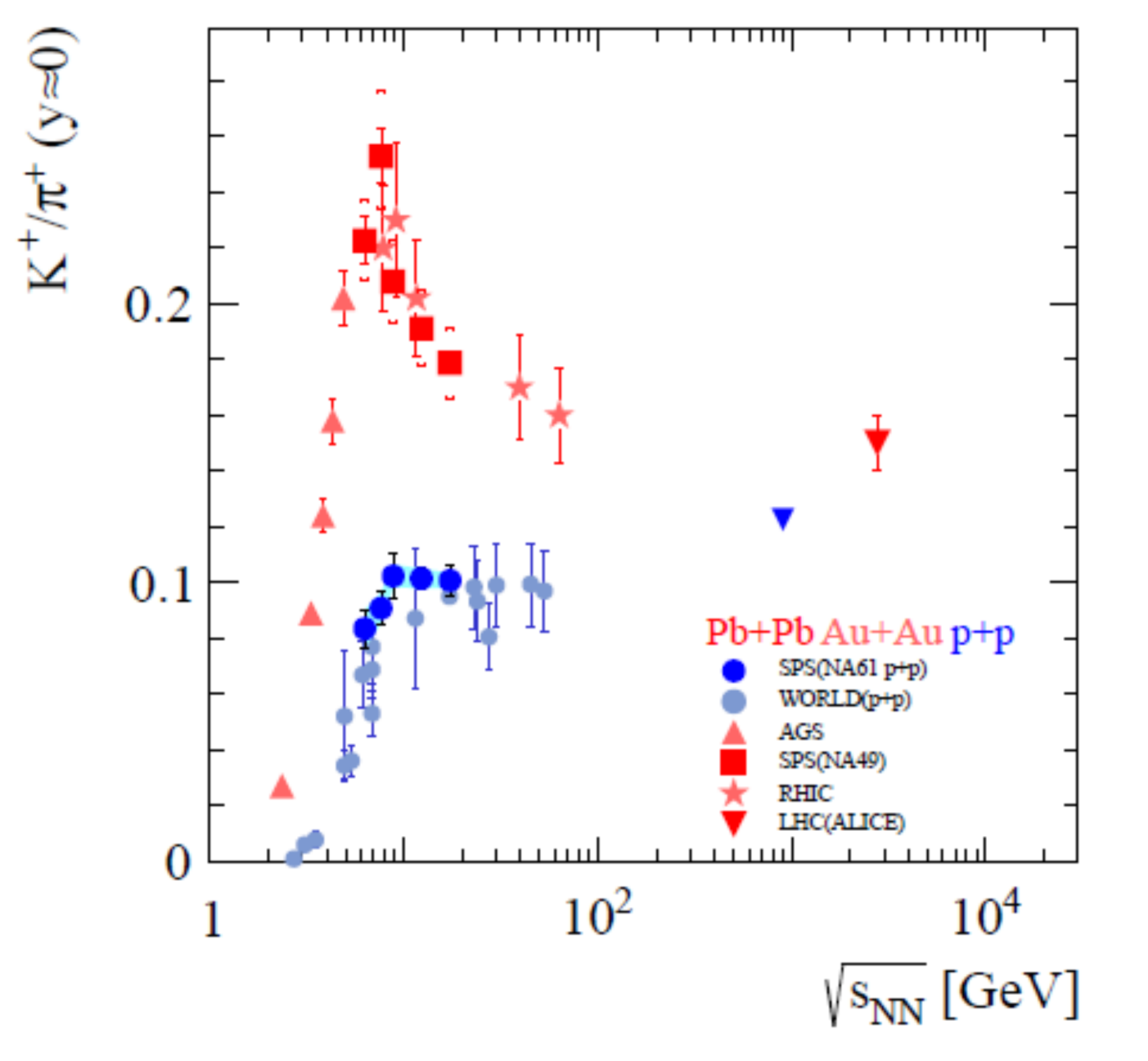}
\caption{Data taking status of the strong interaction program of NA61/SHINE (left). Energy dependence of the ratio of kaon to pion yield in p+p and central Pb+Pb interactions - the "horn" (right).}
\label{plans}
\end{figure}

The broad experimental program of NA61/SHINE includes the study of strong interactions which mainly focuses on the onset of deconfinement (OD) and search for the critical point (CP). The available data suggest that both phenomena may be located in a region of the phase diagram accessible at SPS energy. In order to study properties of the onset of deconfinement and search for the CP NA61/SHINE performs for the first time a two dimensional scan in system size and energy.  Figure~\ref{plans}~(left) shows for which systems and energies data has already been collected (green), is scheduled for recording (red) or is planned (gray). 

\section{Single particle spectra}
\begin{figure}[h]
\includegraphics[width=0.45\textwidth]{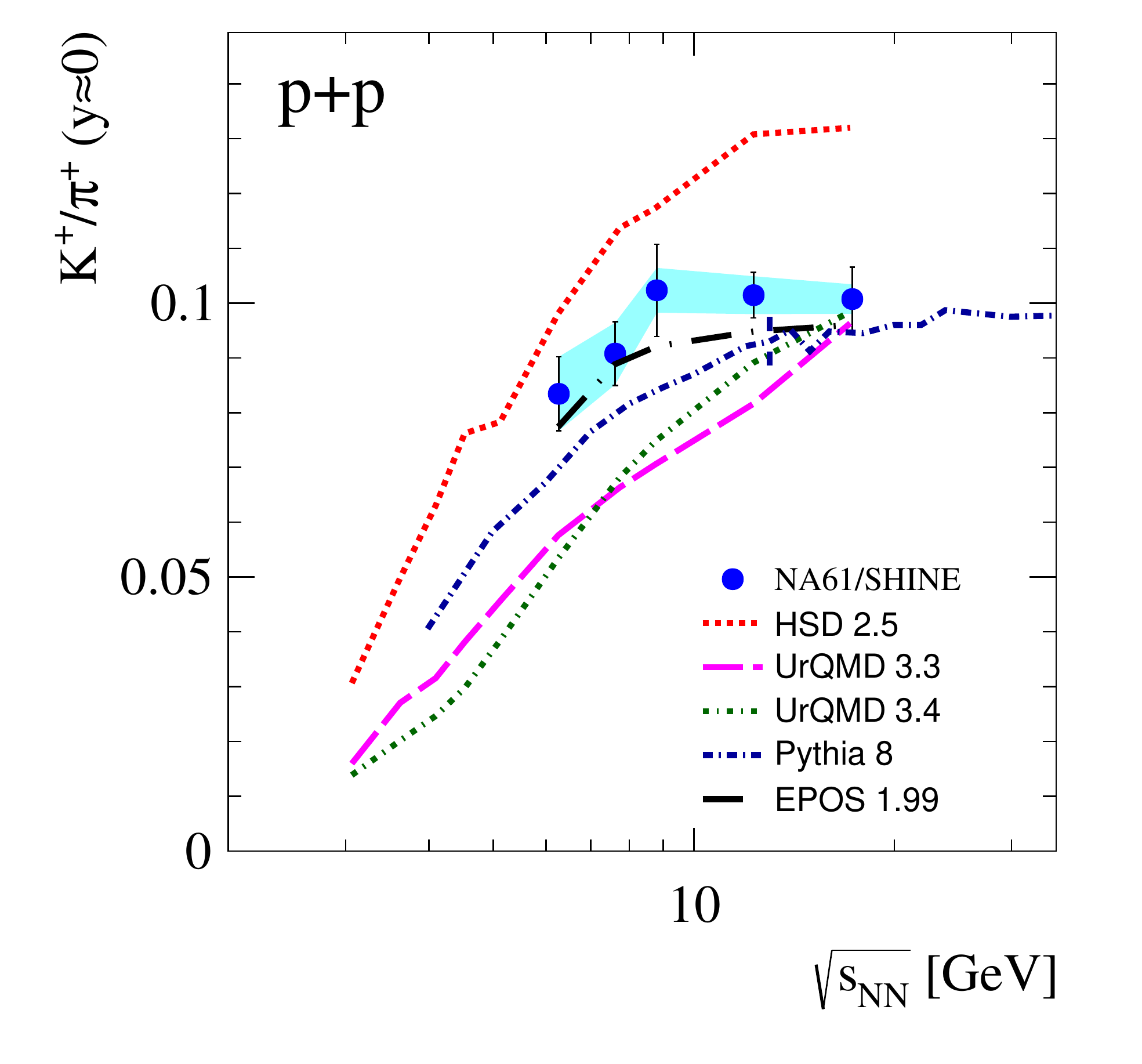}
\includegraphics[width=0.45\textwidth]{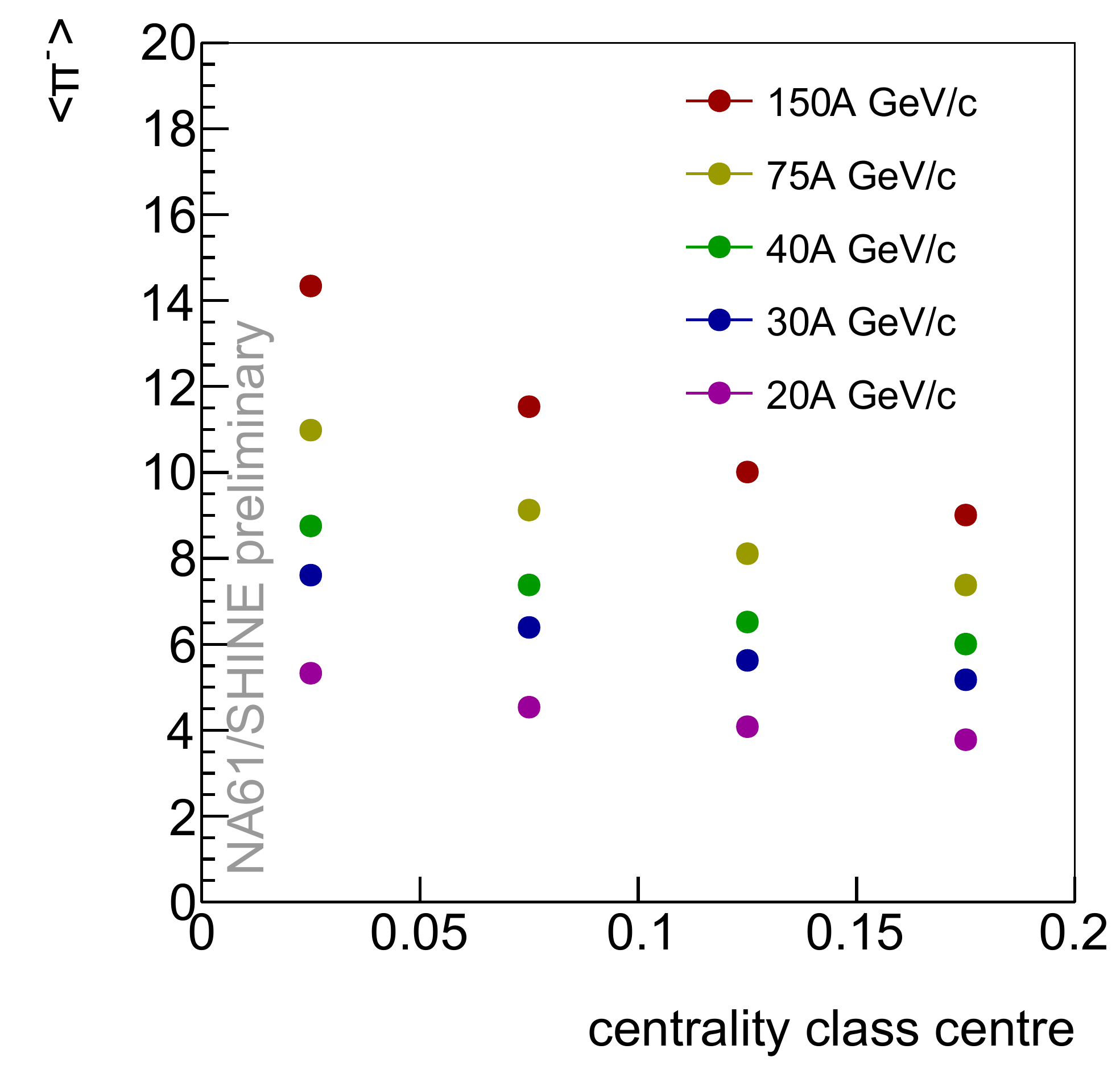}
\label{fig2}
\caption{Comparison of the energy dependence of the ratio of kaon to pion yield in p+p collisions compared to various models (left). Mean multiplicity of negatively charged pions in $^{7}$Be+$^{9}$Be reactions for various collision energies and centalities(right).}
\end{figure}

The Statistical Model of the Early Stage (SMES)~\cite{Gazdzicki:1998vd} suggests that the phase transition line is crossed by heavy ion collisions between the top AGS energy (beam energy 11.7A GeV) and the top SPS energy (beam energy 158A GeV). The SMES model predicts several signatures of the OD. One of them, the "horn", is expected in the strangeness-to-entropy ratio. The ratio should rise with energy in the hadron phase, attain a maximum at the phase transition, and then decrease to an almost constant value. This behavior was experimentally confirmed in central Pb+Pb collisions by the NA49 experiment. In p+p interactions available data did not suggest any structure at these energies. The phase transition region, as seen in NA49, is located between 30A and 40A GeV. The energy dependence of the ratio of kaon to pion yield at mid-rapidity is shown in Fig.~\ref{plans}(right). The much higher statistics collected by NA61/SHINE for p+p interactions as well as good particle identification allowed to improve the quality of the available data. Surprisingly, the results from inelastic p+p interactions exhibit rapid changes (a step-like bahaviour) which looks like a precursor of the "horn" observed  in  central  Pb+Pb collisions. This finding can be qualitatively explained by the SMES model if a phase transition is assumed in p+p interactions and exact strangeness conservations is taken into account which should be included as the number of strangeness carriers is small~\cite{Poberezhnyuk:2015wea}. The precision of the NA61/SHINE data also allows to impose rigorous constraints on Monte-Carlo models~\cite{Pulawski:2015tka}. Comparison with various string hadronic models is shown in Fig.~\ref{fig2} (left).

\begin{figure}[h]
\includegraphics[width=0.45\textwidth]{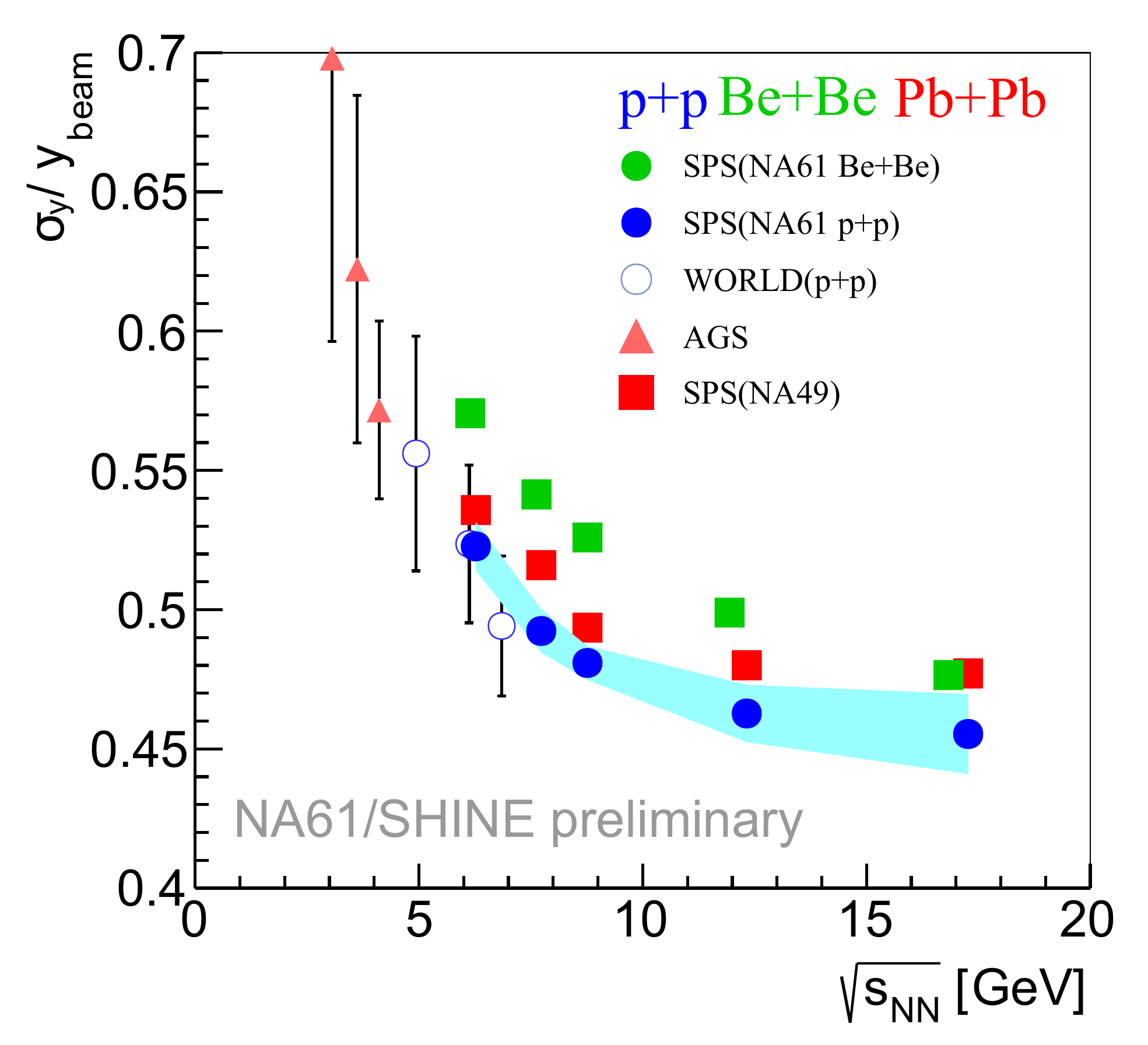}\includegraphics[width=0.45\textwidth]{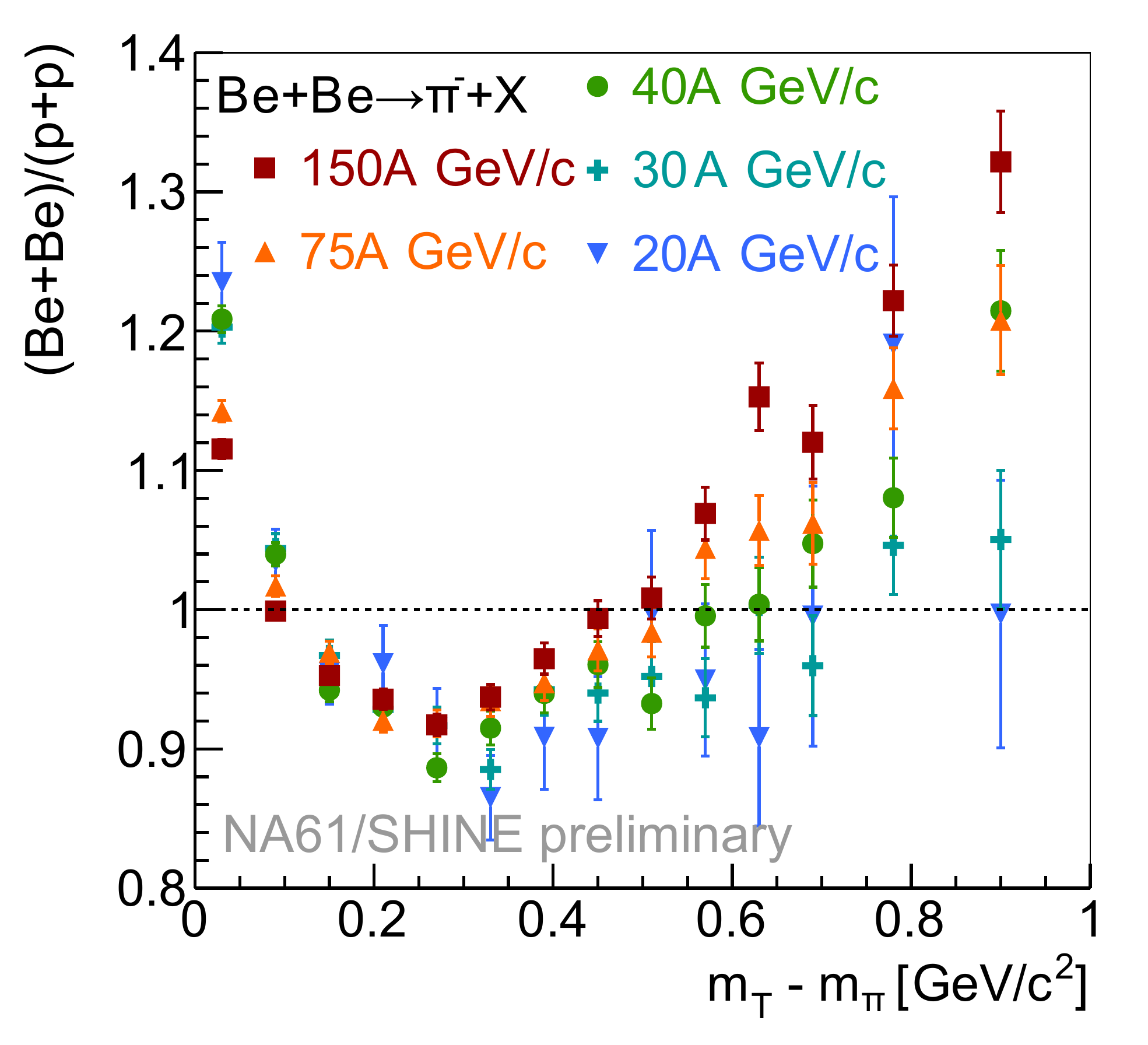}
\caption{Energy dependence of the scaled width $\sigma_{y}/y_{beam}$ of the rapidity distributions (left) and ratio of transverse mass spectra in $^{7}$Be+$^{9}$Be and p+p interactions (right).}
\label{isospin}
\end{figure}

Preliminary results on spectra were obtained for $^{7}$Be+$^{9}$Be interactions at beam momenta of 20-150A GeV/c in the centrality classes 0-5$\%$, 5-10$\%$, 10-15$\%$ and 15-20$\%$~\cite{Kaptur:cpod14}.  Unless otherwise stated, only statistical errors are shown.  The centrality was derived from  the  energy  deposited  in  the  forward  calorimeter, the Particle Spectator Detector. Figure~\ref{fig2} (right) shows the negatively charged pion multiplicity at five beam momenta as a function of centrality class. 
The rapidity spectra (not shown) were fitted with two Gaussians symmetrically displaced with respect to mid-rapidity. The RMS width, $\sigma_{y}$, of the spectra was obtained from the fitted function.  Figure~\ref{isospin} (left) shows the scaled width $\sigma_{y}/y_{beam}$ for $^{7}$Be+$^{9}$Be as well as inelastic p+p and central Pb+Pb interactions.  The widths  of  the  spectra  for  all  systems  decrease  monotonically  with  respect  to  collision  energy. However,  the  widths  of  the  spectra  do  not  behave  monotonically  with  respect  to  the  system size for a given collision energy.  The widths of the spectra from Pb+Pb collisions lie between those of p+p and $^{7}$Be+$^{9}$Be interactions.  When comparing the rapidity spectra from different systems, one must consider that p+p system has larger isospin asymmetry than $^{7}$Be+$^{9}$Be and Pb+Pb.  

Measured transverse mass spectra of $\pi^{-}$ in $^{7}$Be+$^{9}$Be are approximately exponential (for details see Ref.~\cite{Kaptur:cpod14}). The ratio of $^{7}$Be+$^{9}$Be to p+p interactions  is shown in Fig.~\ref{isospin}~(right). The observed energy dependence of the ratio may indicate the presence of collective radial flow in $^{7}$Be+$^{9}$Be interactions. 
\section{Fluctuations and correlations}
Observation of the onset of deconfinement at lower SPS energies suggests the possibility of detecting the CP in heavy ion collisions.A specific property of the CP - the increase in the correlation length - makes fluctuations its basic signal. In order to observe the CP, created matter should freeze-out near its location. So, if at all, the CP signals are expected at energies higher than those of the onset of deconfinement. Indications of enlarged fluctuations in multiplicity and transverse momentum as well as an intermittency signal were found by NA49 in Si+Si collisions at 158A GeV~\cite{KG,Anticic:2012xb}. 
	
When comparing fluctuations in systems of different size (volume) one should use quantities which are insensitive to volume and volume fluctuations (unavoidable in A+A collisions). Two families of such quantities were used. First, the intensive quantity scaled variance, $\omega[A] = \frac{\langle A^{2}\rangle-\langle A\rangle^{2}}{\langle A \rangle}$ where $A$ stands for an event quantity. In the Grand Canonical Ensemble it is independent of volume but it depends on volume fluctuations. Second, the strongly intensive quantities $\Sigma[A,B]$ and $\Delta[A,B]$ defined as:
\begin{equation}
	\Delta[A,B] = \frac{1}{C_{\Delta}}\big[ \langle B \rangle\omega_{A} - \langle A \rangle\omega_{B} \big], \quad
	\Sigma[A,B] = \frac{1}{C_{\Sigma}}\Big[ \langle B \rangle\omega_{A} + \langle A \rangle\omega_{B} - 2 \big( \langle AB \rangle - \langle A \rangle \langle B \rangle \big) \Big], 
\end{equation}
where $A$ and $B$ are two different event quantities. For comparison with the NA49 experiment also the $\Phi$ quantity was used. For fluctuations of $P_{T}=\sum\limits_{k=1}^{N} p_{T_{k}}$ it reads $\Phi_{p_{T}}=\sqrt{\overline{p_{T}}\omega[p_{T}]} \big[ \sqrt{\Sigma[P_{T},N]}-1 \big]$. For fluctuations of multiplicities of two types of particles $N_i$, $N_j$ it is defined as $\Phi_{ij}=\frac{\sqrt{\langle N_{i}\rangle\langle N_{j}\rangle}}{\langle N_{i}+N_{j}\rangle}\cdot[\sqrt{\Sigma[i,j]}-1]$.  Strongly intensive quantities do not depend on volume and volume fluctuations but one needs two event quantities to construct them. 

\begin{figure}
\includegraphics[width=0.45\textwidth]{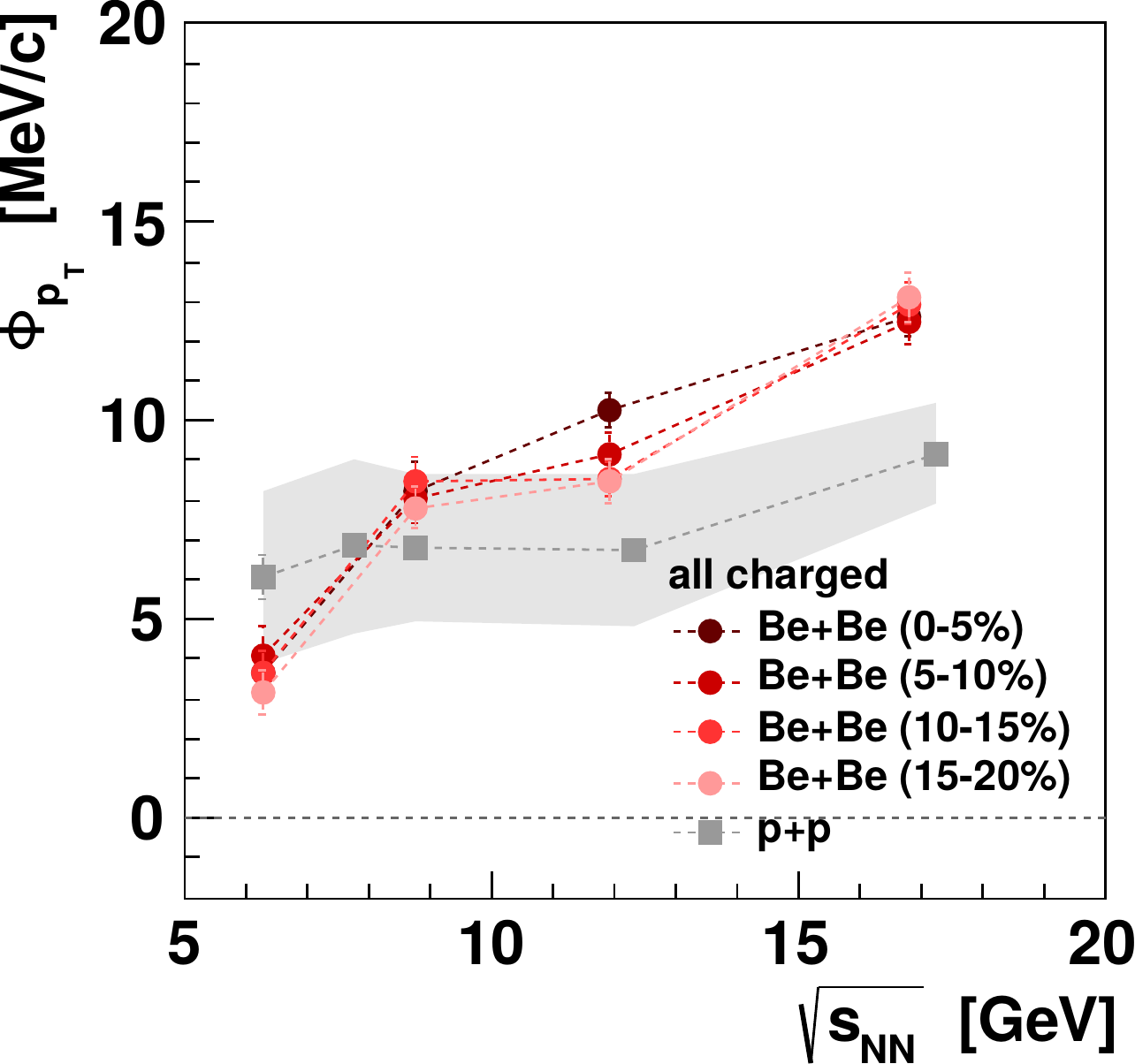}
\includegraphics[width=0.45\textwidth]{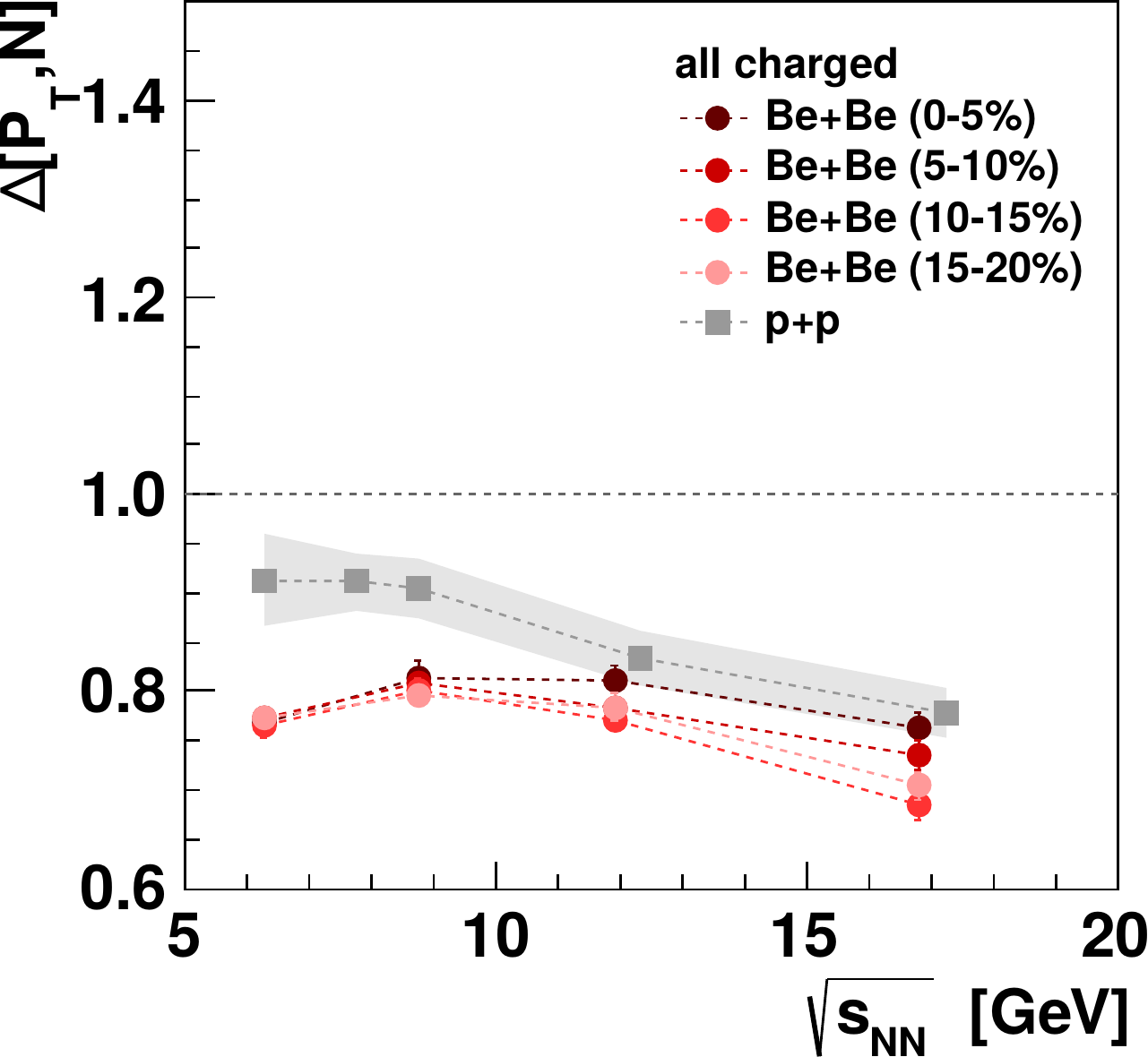}
\label{siq}
\caption{The energy dependence of $\Sigma[P_{T},N]$ (left) and $\Delta[P_{T},N]$ (right) in p+p and $^{7}$Be+$^{9}$Be interactions.}
\end{figure}

Enhanced fluctuations of multiplicity and mean transverse momentum were suggested as a possible signature of the CP of strongly interacting matter~\cite{Stephanov:1999zu}.  Transverse momentum fluctuations were measured by the NA49 experiment using the $\Phi_{p_{T}}$ quantity~\cite{phiref}. For system size dependence at the top SPS energy NA49 finds a maximum of $\Phi_{p_{T}}$ for C+C and Si+Si collisions~\cite{KG}. The energy dependence of $\Phi_{P_{T}}$ and $\Delta[P_{T},N]$ in p+p and $^{7}$Be+$^{9}$Be interactions is shown in Fig.~\ref{siq}. The values in $^{7}$Be+$^{9}$Be interactions exhibit no centrality dependence  and are close to p+p results. No structures which could be connected to the CP/OD in p+p and $^{7}$Be+$^{9}$Be collisions are visible. Bose-Einstein statistics and $\langle p_{T} \rangle$, $N$ correlations probably introduce differences between $\Sigma[P_{T},N]$ and $\Delta[P_{T},N]$. Details on this analysis can be found in Ref.~\cite{::2015jna}.

\begin{figure}
\includegraphics[width=0.45\textwidth]{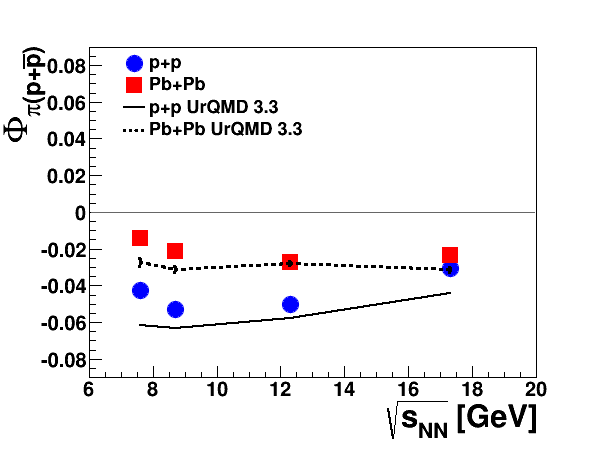}
\includegraphics[width=0.45\textwidth]{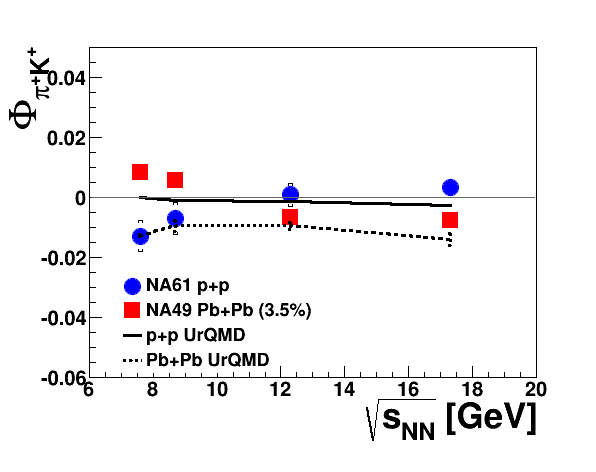}
\label{phi}
\caption{The energy dependence of $\Phi_{\pi(p+\bar{p})}$ (left) and $\Phi_{\pi^{+}K^{+}}$ (right) in p+p (NA61/SHINE) and central Pb+Pb (NA49) interactions~\cite{Rustamov:2013oza}.}
\end{figure}

Several mechanisms could lead to specific event-by-event particle ratio (chemical) fluctuations. Among them are overheating-supercooling fluctuations due to a first order phase transition with considerable latent heat, or fluctuations due to coexistence of confined and deconfined matter (mixed phase)~\cite{chem}. Chemical fluctuations might contain otherwise inaccessible information
about the nature and order of the phase transition at these energies~\cite{chem}. They are also
influenced by conservation laws, quantum statistics, and resonance decays. Comparison with Pb+Pb interactions from NA49 is shown in Fig.~\ref{phi}. The results for $\Phi_{\pi(p+\bar{p})}$ and $\Phi_{\pi^{+}K^{+}}$ in both reactions are similar. $\Phi_{\pi(p+\bar{p})}$ is probably dominated by conservation laws and resonance decays (agreement with UrQMD model supports this interpretation). No structures can be connected with the CP but there is a systematic difference in $\Phi_{\pi^{+}K^{+}}$ possibly connected to OD. Further studies are needed in order to understand the observed discrepancy.

\begin{figure}
\begin{center}
\includegraphics[width=1.8in]{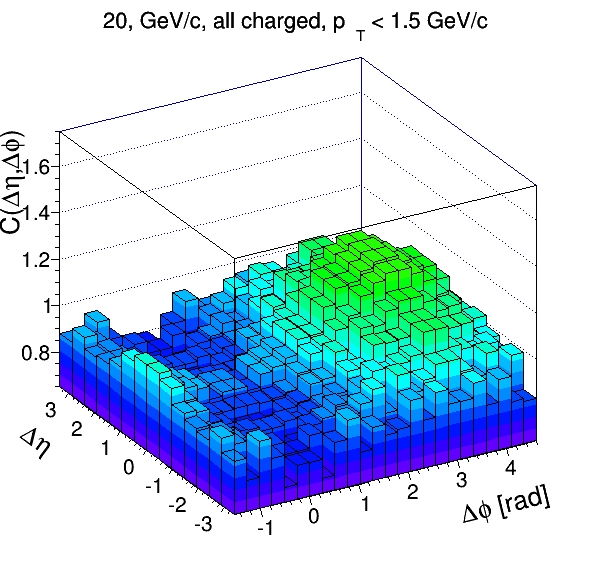}
\includegraphics[width=1.8in]{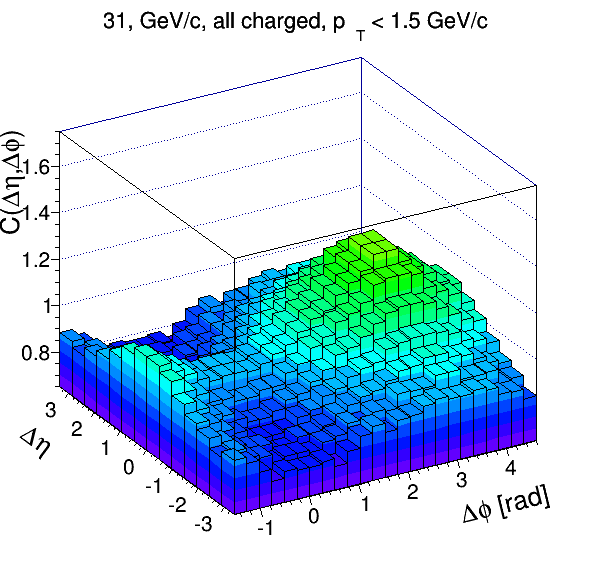}
\includegraphics[width=1.8in]{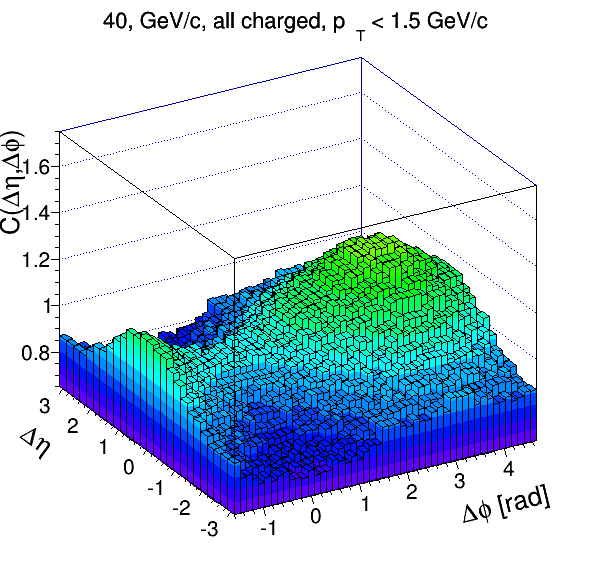}\newline
\includegraphics[width=1.8in]{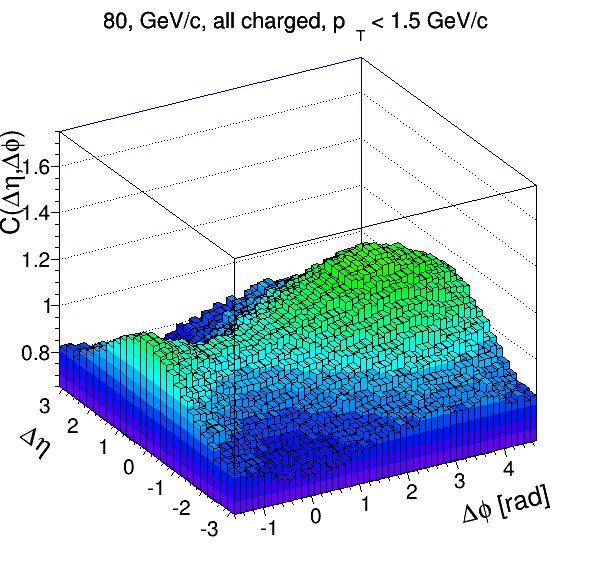}
\includegraphics[width=1.8in]{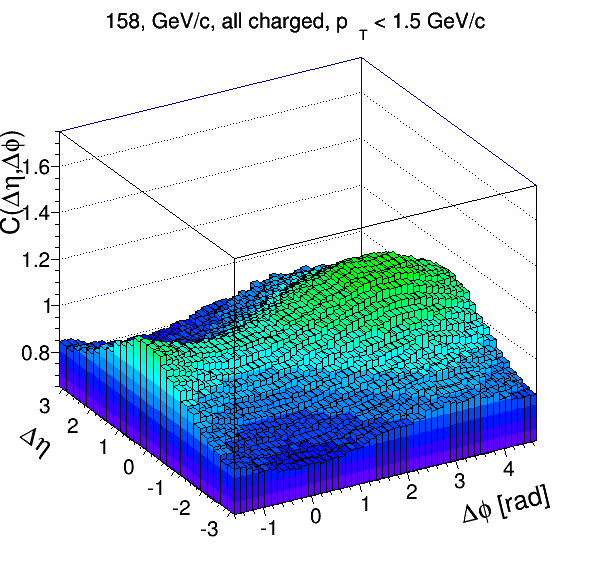}
\end{center}
\label{deltaphi}
\caption{Correlation function $C(\Delta\eta\Delta\phi)$ in inelastic p+p interactions for all charged particle pairs. The correlation function is mirrored around $(\Delta\eta,\Delta\phi) = (0,0)$.}
\end{figure}

Two-particle correlations in $\Delta\eta\Delta\phi$ were studied extensively at RHIC and LHC. They may allow to disentangle different sources of correlations: jets, flow, resonance decays, quantum statistics effects, conservation laws, etc. Correlations are calculated as a function of the difference in pseudo-rapidity ($\eta$) and azimuthal angle ($\phi$) between two particles in the same event. The correlation function $C(\Delta\eta,\Delta\phi)$ is obtained as the ratio of distributions of pairs from data and mixed events of the same multiplicity, respectively. Details can be found in Ref.~\cite{Bartek}. The energy dependence of the correlation functions from p+p reactions for all charged pair combinations is presented in Fig.~\ref{deltaphi} for different collision energies.
Several structures can be seen in the plots:
\begin{itemize}
	\item A maximum at $(\Delta\eta,\Delta\phi)=(0,\pi)$,  probably a result of resonance decays and momentum conservation. It is strongest for unlike-sign pairs and significantly weaker for same charge pairs
	\item A weak enhancement at $(\Delta\eta,\Delta\phi)=(0,0)$,  likely due to Coulomb interactions (unlike-sign pairs) and quantum statistics (same charge pairs).
\end{itemize}
\section{Summary}
The analysis of inelastic p+p and $^{7}$Be+$^{9}$Be interactions at CERN SPS energies showed many
interesting  effects. In particular, the p+p data exhibited step-like structures in the energy
region where the onset of deconfinement was found in central Pb+Pb collisions.  Also, precise NA61/SHINE measurements can not be explained well by theoretical models. For $^{7}$Be+$^{9}$Be reactions an indication of a transverse flow effect was found at the highest beam momenta. There are no indications of the Critical Point in fluctuations of transverse momentum in p+p and $^{7}$Be+$^{9}$Be interactions. Fluctuations and correlations in p+p and $^{7}$Be+$^{9}$Be are dominated by conservation laws and resonance decays.\newline

This work was supported by
the Hungarian Scientific Research Fund (grants OTKA 68506 and 71989),
the J\'anos Bolyai Research Scholarship of
the Hungarian Academy of Sciences,
the Polish Ministry of Science and Higher Education (grants 667\slash N-CERN\slash2010\slash0, NN\,202\,48\,4339 and NN\,202\,23\,1837),
the Polish National Center for Science (grants~2011\slash03\slash N\slash ST2\slash03691, 2012\slash04\slash M\slash ST2\slash00816,
2013\slash11\slash N\slash ST2\slash03879, 2014\slash 13\slash N\slash ST2\slash 02565),
the Foundation for Polish Science --- MPD program, co-financed by the European Union within the European Regional Development Fund,
the Federal Agency of Education of the Ministry of Education and Science of the
Russian Federation (SPbSU research grant 11.38.193.2014),
the Russian Academy of Science and the Russian Foundation for Basic Research (grants 08-02-00018, 09-02-00664 and 12-02-91503-CERN),
the Ministry of Education, Culture, Sports, Science and Tech\-no\-lo\-gy, Japan, Grant-in-Aid for Sci\-en\-ti\-fic Research (grants 18071005, 19034011, 19740162, 20740160 and 20039012),
the German Research Foundation (grant GA\,1480/2-2),
the EU-funded Marie Curie Outgoing Fellowship,
Grant PIOF-GA-2013-624803,
the Bulgarian Nuclear Regulatory Agency and the Joint Institute for
Nuclear Research, Dubna (bilateral contract No. 4418-1-15\slash 17),
Ministry of Education and Science of the Republic of Serbia (grant OI171002),
Swiss Nationalfonds Foundation (grant 200020\-117913/1)
and ETH Research Grant TH-01\,07-3.

\end{document}